\documentclass{emulateapj}
\usepackage{graphicx,graphics,amsmath}

\renewcommand{\vec}[1]{\mbox{\boldmath $#1$}}

\shorttitle{Baroclinic Instability in Stars}
\shortauthors{L.\,L.~Kitchatinov}

\begin{document}

\title{Baroclinic Instability in Stellar Radiation Zones}

\author{L.\,L.~Kitchatinov$^{1,2}$}
\affil{\it $^1$Institute for Solar-Terrestrial Physics, Lermontov Str. 126A, Irkutsk 664033, Russia
\\
$^2$Pulkovo Astronomical Observatory, St. Petersburg 176140, Russia}
\email{kit@iszf.irk.ru}


\begin{abstract}
Surfaces of constant pressure and constant density do not coincide in differentially rotating stars. Stellar radiation zones with baroclinic stratification can be unstable.
Instabilities in radiation zones are of crucial importance for angular momentum transport, mixing of chemical species and, possibly, for magnetic field generation. This paper performs linear analysis of baroclinic instability in differentially rotating stars.
Linear stability equations are formulated for differential rotation of arbitrary shape and then solved numerically for rotation non-uniform in radius.
As the differential rotation increases, $r$- and $g$-modes of initially stable global oscillations transform smoothly into growing modes of baroclinic instability. The instability can therefore be interpreted as stability loss to $r$- and $g$-modes excitation. Regions of stellar parameters where $r$- or $g$-modes are preferentially excited are defined. Baroclinic instability onsets at a very small differential rotation of below 1\%. The characteristic time of instability growth is about one thousand rotation periods. Growing disturbances possess kinetic helicity. Magnetic field generation by the turbulence resulting from baroclinic instability in differentially rotating radiation zones is, therefore, possible.
\end{abstract}

\section{INTRODUCTION}
Stratification in not too rapidly rotating stars is close to spherical symmetry. Disregarding deviations from this symmetry, only two possibilities for mutual orientation of the gradients of pressure and entropy are possible. Both possibilities of parallel and antiparallel orientation  are realised in the convection and radiation zones of stars respectively. Stratification in radiation zones is believed to be stable because radial displacements are opposed by buoyancy. The stabilizing effect of subadiabatic stratification is, however, obvious for strictly parallel gradients of entropy and pressure only. Even a slight deviation from barotropic stratification can provoke an instability (\citeauthor{T00} \citeyear{T00}, chapter 3).

\begin{figure}[htb]
    \begin{center}
    \includegraphics[width=0.45\textwidth]{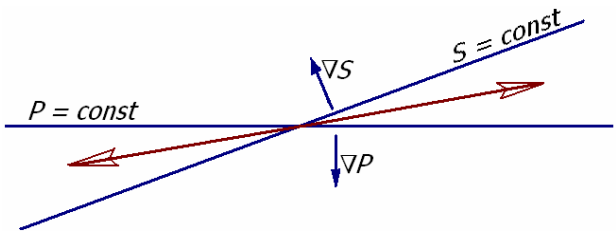}
    \end{center}
    \caption{If the surfaces of constant entropy and constant pressure do not coincide, adiabatic displacements in a narrow cone between these surfaces (long arrows) are supported by buoyancy.
    }
    \label{f1}
\end{figure}

Figure\,\ref{f1} illustrates the origin of instability expected for radiation zones with baroclinic stratification. Displacements in narrow cones between isentropic and isobaric surfaces are potentially unstable. Projections of the displacements in the directions of the entropy and pressure gradients have the same sign. Therefore, the buoyancy force supports the displacements. Instability can develop at the  expense of gravitational energy \citep{S80}.

Baroclinicity, in turn, naturally results from non-uniform rotation. If angular velocity varies with the cylindrical coordinate $z$ along the rotation axis, the centrifugal force is not conservative so that the pressure force per unit mass, $\rho^{-1}{\vec\nabla}P$, should not be conservative either, ${\vec\nabla}\rho\times{\vec\nabla}P \neq 0$, in order to balance the centrifugal force. Baroclinic instability can be expected for rotating radiation zones with $z$-dependent angular velocity.

Studies of baroclinic instability in an astrophysical context has a long history. Already \citet{GS67} found the condition $\mathrm{d}\Omega/\mathrm{d}z = 0$ as necessary for the stability of rotating radiation zones. \citet{S80} noticed the significance of the relation between differential rotation and baroclinicity for stability. These and related studies \citep{A78,SK84,K91} considered local stability. The spatial scale of disturbances was assumed to be small compared to the scale height. This paper concerns stability against disturbances that are global in horizontal dimensions. The radial scale of disturbances is still assumed to be short. Subadiabatic stratification in radiation zones precludes mixing on large radial scales. A similar approach was applied to (barotropic) instabilities in the solar tachocline \citep{Cea99,Gea07,KR09} to find that the dominating modes of the instabilities are indeed global in horizontal dimensions. We shall see that the same is true of baroclinic instability. It was suggested in the preceding Letter (\citeauthor{K13}\,\citeyear{K13}; K13 hereafter) that growing modes of baroclinic instability correspond to global Rossby waves ($r$-modes) and internal gravity waves ($g$-modes). This paper confirms this expectation by following the dependence of eigenmodes on differential rotation. As the differential rotation and the resulting baroclinicity increase from zero, $r$- and $g$-modes transform smoothly into unstable eigenmodes. Baroclinic instability can, therefore, be interpreted as stability loss to excitation of $r$- and $g$-modes. The regions in parametric space, where a specific mode dominates, will be identified.

Differential rotation is known to affect gravito-inertial waves in an essential way \citep{A85,LS93,M09,BR13,AMD13}. We consider this influence taking into account the deviation of stratification from barotropy and find that some of the modified $r$- and $g$-modes are amplified while some others are damped due to baroclinicity induced by the differential rotation.

Unstable modes posses kinetic helicity, ${\vec u}\cdot ({\vec\nabla}\times{\vec u}) \neq 0$. Helicity is important for generation of global magnetic fields
(cf. \citet{M78} or a more recent discussion of helicity effects by \citet{BS05}).
The instability can, therefore, be relevant to the origin of magnetic fields of stellar radiation zones.

Section~\ref{LSP} provides mathematical formulation of the problem. Linear stability equations are formulated for differential rotation of arbitrary shape. Section~\ref{RD} presents and discusses the results. The earlier guess (K13) that the growing modes of baroclinic instability can be understood as  modified $r$- and $g$-modes of stable oscillations is confirmed by computations. Detailed stability map showing the regions in parametric space where $r$- or $g$-modes are preferentially excited is constructed. Computations show that unstable modes possess kinetic helicity. Helicity patterns for $r$- and $g$-modes are compared. Section~\ref{Con} summarises the results and discusses their implications. It is noted in particular that toroidal magnetic field induces baroclinicity and may lead to baroclinic instability similar to the differential rotation.
\section{LINEAR STABILITY PROBLEM}\label{LSP}
\subsection{Background Equilibrium}
The background equilibrium is assumed to be axially symmetric about the rotation axis. Hydrodynamic stability is considered, i.e. the magnetic field is neglected. The steady motion equation then reads
\begin{equation}
    ({\vec V}\cdot{\vec\nabla}){\vec V} = -\frac{1}{\rho}{\vec\nabla}P - {\vec\nabla}\psi ,
    \label{1}
\end{equation}
where $\psi$ is the gravity potential, other notations are standard, and the effect of viscosity on the global flow is neglected. The principal motion in the radiation zone is rotation,
\begin{equation}
    {\vec V} = {\vec e}_\phi r \sin\theta \Omega .
    \label{2}
\end{equation}
In this equation, the usual spherical coordinates $(r,\theta ,\phi )$ are used, ${\vec e}_\phi$ is the azimuthal unit vector, and $\Omega$ is the angular velocity.

The meridional flow in subadiabatically stratified radiation zones is small. The characteristic time of meridional circulation in the solar radiation zone exceeds the age of the sun \citep{T00}. Nevertheless, the meridional motion equation gives important condition of balance of meridional forces,
\begin{equation}
    r \sin\theta\frac{\partial\Omega^2}{\partial z} =
    -\frac{1}{\rho^2}\left({\vec\nabla}\rho\times{\vec\nabla}P\right)_\phi,
    \label{3}
\end{equation}
where $\partial /\partial z = \cos\theta\partial /\partial r - r^{-1}\sin\theta\partial /\partial\theta$ is gradient along the rotation axis. Condition (\ref{3}) can be obtained by taking the azimuthal component of curled equation (\ref{1}). Centrifugal force in  rotational motion is conservative only if the angular velocity does not vary with the cylindrical coordinate $z$. The left side of Eq.\,(\ref{3}) accounts for the non-conservative part of centrifugal force. This force alone would drive a vortical meridional flow. In the radiation zone of a star, however, the force is balanced by buoyancy accounted for by the right side of Eq.\,(\ref{3}). The equilibrium is baroclinic: the surfaces of constant pressure and density do not coincide. The surfaces of constant pressure and entropy $s = c_\mathrm{v}\ln (P) - c_\mathrm{p} \ln (\rho )$ differ as well:
\begin{equation}
    {\vec\nabla}\rho\times{\vec\nabla}P = -\frac{\rho}{c_\mathrm{p}}{\vec\nabla}s\times{\vec\nabla}P = -\frac{\rho^2}{c_\mathrm{p}}{\vec\nabla}s\times{\vec g}^* ,
    \label{4}
\end{equation}
where ${\vec g}^* = -{\vec\nabla}\psi + r\sin\theta\Omega ({\vec e}_\phi\times{\vec\Omega})$ is the effective gravity. The instability considered in this paper is caused by the baroclinicity of the stratification, not by the rotationally induced deviation from spherical symmetry. We neglect the deviation of isobaric surfaces from spheres but keep the baroclinicity as it is defined by Eq.\,(\ref{4}),
\begin{equation}
    \frac{\partial s}{\partial\theta} = 2g^{-1}c_\mathrm{p}r\sin\theta\Omega
    \left( \cos\theta\ r\frac{\partial\Omega}{\partial r}
    - \sin\theta\frac{\partial\Omega}{\partial\theta}\right) ,
    \label{5}
\end{equation}
where $g$ is gravity. For the case of shellular rotation where $\Omega$ depends on $r$ only, the expression for the meridional gradient of entropy simplifies further to read
\begin{equation}
    \frac{\partial s}{\partial\theta} =
    -2qg^{-1}c_\mathrm{p}r\Omega^2 \sin\theta\cos\theta ,
    \label{6}
\end{equation}
where
\begin{equation}
    q = -\frac{r}{\Omega}\frac{\mathrm{d}\Omega}{\mathrm{d}r}
    \label{7}
\end{equation}
is the normalized shear of the radial differential rotation.
\subsection{Linear Stability Equations}
The derivation of the linear stability equations in this paper is almost identical to that of earlier studies \citep{KR08,K08}. The difference is only that baroclinicity is now included and magnetic fields are neglected. The eigenvalue equations will, therefore, be written without repeating their derivations.
The equations of this paper generalize those of K13 by allowing for differential rotation of arbitrary shape, not only in radius. We recap now the main assumptions and approximations used in the derivations.

The background equilibrium does not depend on time or longitude. Dependencies of linear disturbances on time and azimuth can, therefore, be assumed to be exponential, $\mathrm{exp}\left(\mathrm{i}m\phi - \mathrm{i}\omega t\right)$, where $m$ is the azimuthal wave number. A positive imaginary part in the eigenvalue, $\Im (\omega ) > 0$, means an instability.

Subadiabatic stratification in the radiation zone precludes mixing on large radial scales. The radial scale of disturbances is, therefore, assumed to be small and the stability analysis is local in radius. This means that the disturbances of velocity $u$ and entropy $s'$ depend on radius as  $\mathrm{exp}\left(\mathrm{i}kr\right)$ and $kr \gg 1$. Horizontal mixing on the contrary is not opposed by buoyancy and the stability analysis is global in horizontal dimensions. We shall see that the most rapidly growing modes actually have large horizontal scales.

Non-compressive disturbances are considered, $\mathrm{div}{\vec u} = 0$. This assumption can be justified for disturbances with the short radial wave length compared with the density scale height. The density disturbances in the buoyancy terms are, however, retained. This is very similar to the standard Boussinesq approximation with the only difference that entropy not temperature is conserved by adiabatic displacements in the (gaseous) radiation zone.

The flow equations are formulated in terms of the scalar potentials, $P_u$ and $T_u$, for the poloidal and toroidal parts, respectively, of the velocity field,
\begin{eqnarray}
    {\vec u} = \frac{{\vec e}_r}{r^2}\hat{L}P_u
    &-& \frac{{\vec e}_\theta}{r}\left(\frac{\mathrm{i}m}{\sin\theta} T_u + \mathrm{i}k \frac{\partial P_u}{\partial\theta}\right)
    \nonumber \\
    &+& \frac{{\vec e}_\phi}{r}\left(\frac{\partial T_u}{\partial\theta} + \frac{k m}{\sin\theta}P_u\right)
    \label{8}
\end{eqnarray}
\citep{C61}, where
\begin{equation}
    \hat{L} = \frac{1}{\sin\theta}\frac{\partial}{\partial\theta} \sin\theta\frac{\partial}{\partial\theta} + \frac{1}{\sin^2\theta}\frac{\partial^2}{\partial\phi^2}
    \label{9}
\end{equation}
is the angular part of the Laplacian operator. Dimensionless variables are used. Physical variables can be restored from normalized entropy ($S$), toroidal ($W$) and poloidal ($V$) flow potentials by using Eq.\,(\ref{8}) and following relations
\begin{equation}
    s' = -\frac{\mathrm{i}c_\mathrm{p}N^2}{gk} S,\ \
    P_u = \left(\Omega_0 r^2/k\right) V,\ \
    T_u = \Omega_0 r^2 W ,
    \label{10}
\end{equation}
where $\Omega_0$ is the characteristic value of angular velocity.

The mathematical formulation of the stability problem reduces to the eigenvalue problem for a set of three ordinary differential equations with latitude as the independent variable. The equations for toroidal flow,
\begin{eqnarray}
    (\hat{\omega} &-& m \hat{\Omega} ) \left(\hat{L}W\right)\ =
    -\mathrm{i}\frac{\epsilon_\nu}{\hat{\lambda}^2} \left(\hat{L}W\right)
    \nonumber \\
     -&m& \frac{\partial^2\left((1-\mu^2)\hat{\Omega}\right)}{\partial\mu^2}\ W
    + \frac{\partial\left((1-\mu^2)\hat{\Omega}\right)}{\partial\mu}
    \left(\hat{L}V\right)
    \nonumber \\
    &+& \frac{\partial^2\left((1-\mu^2)\hat{\Omega}\right)}{\partial\mu^2}\
    (1-\mu^2)\frac{\partial V}{\partial\mu} ,
    \label{11}
\end{eqnarray}
and poloidal flow,
\begin{eqnarray}
    (\hat{\omega} &-& m \hat{\Omega}) \left(\hat{L}V\right)\ =
    -\mathrm{i}\frac{\epsilon_\nu}{\hat{\lambda}^2} \left(\hat{L}V\right) -
    \hat{\lambda}^2 \left(\hat{L}S\right)
    \nonumber \\
    &+& 2 m\left(\frac{\partial (\mu\hat{\Omega})}{\partial\mu}\ V + (1 -
    \mu^2)\frac{\partial\hat{\Omega}}{\partial\mu}\frac{\partial
    V}{\partial\mu}\right)
    \\
    \label{12}
    &-& 2\mu\hat{\Omega}\left(\hat{L}W\right) - 2
    (1-\mu^2)\frac{\partial(\mu\hat{\Omega})}{\partial\mu}
    \frac{\partial W}{\partial\mu} - 2 m^2
    \frac{\partial\hat{\Omega}}{\partial\mu}\ W ,
    \nonumber
\end{eqnarray}
are identical to that of the barotropic stability analysis \citep{KR09}. They are not modified by allowance for baroclinicity. In these equations, $\hat\omega = \omega/\Omega_0$ is the normalized eigenvalue, $\hat\Omega = \Omega/\Omega_0$ is the normalized angular velocity, $\mu = \cos\theta$,
\begin{equation}
    \hat\lambda = \frac{N}{\Omega_0 kr}
    \label{13}
\end{equation}
is the key parameter controlling the effect of subadiabatic stratification ($\hat\lambda$ can also be understood as the normalized radial wavelength), and $N$ is the buoyancy frequency
\begin{equation}
    N^2 = \frac{g}{c_\mathrm{p}}\frac{\partial s}{\partial r}.
    \label{14}
\end{equation}
Finite diffusion is included in Eqs. (\ref{11}) and (\ref{12}) via the parameters
\begin{equation}
    \epsilon_\nu = \frac{\nu N^2}{\Omega^3_0 r^2} ,\ \ \
    \epsilon_\chi = \frac{\chi N^2}{\Omega^3_0 r^2} ,
    \label{15}
\end{equation}
where $\nu$ is the viscosity and $\chi$ is the thermal diffusivity.

In the case of baroclinic stratification, entropy disturbances are produced not only by radial displacements but also by meridional motions. This effect is included in the entropy equation,
\begin{eqnarray}
    (\hat{\omega} &-& m \hat{\Omega}) S =
    -\mathrm{i}\frac{\epsilon_\chi}{\hat{\lambda}^2} S
    + \hat{L}V
    \nonumber \\
    &-& \frac{\mathrm{i}\Omega_0}{\hat{\lambda}N}
    \left( \mu r\frac{\partial {\hat\Omega}^2}{\partial r}
    + (1 - \mu^2)\frac{\partial {\hat\Omega}^2}{\partial\mu}\right)
    \nonumber \\
    &&\times\left( mW - (1-\mu^2)\frac{\partial V}{\partial\mu}\right) ,
    \label{16}
\end{eqnarray}
via its second line. Eq.\,(\ref{5}) was used to express baroclinicity in terms of the angular velocity gradient when deriving Eq.\,(\ref{16}). In the case of shellular rotation, $\partial\Omega /\partial\theta = 0$, the entropy equation simplifies to
\begin{eqnarray}
    (\hat{\omega} &-& m \hat{\Omega}) S =
    -\mathrm{i}\frac{\epsilon_\chi}{\hat{\lambda}^2} S
    + \hat{L}V
    \nonumber \\
    &+& \mathrm{i}\frac{Q}{\hat{\lambda}}\mu{\hat\Omega}^2\left( mW - (1-\mu^2)\frac{\partial V}{\partial\mu}\right) ,
    \label{17}
\end{eqnarray}
where
\begin{equation}
    Q = 2 q \frac{\Omega_0}{N} ,
    \label{18}
\end{equation}
and $q$ is the shear parameter of Eq.\,(\ref{7}).

All computations in this paper are performed for the case of shellular rotation. One can put $\hat{\Omega} = 1$ in this case. It was found convenient, however, to leave $\hat\Omega$ as a parameter of the equation system for the following reason. The $\Omega_0$ used for normalization purposes can formally have any value not necessarily equal to the local angular velocity. Then, the dependence on rotation rate can be studied by varying $\hat\Omega$. In this way it will be shown in particular that $g$-modes of oscillations in non-rotating fluid ($\hat\Omega = 0$) transform smoothly into unstable modes of baroclinic instability as the rotation rate increases gradually to $\hat\Omega = 1$.

The stability problem with equations (\ref{11}), (\ref{12}) and (\ref{17}) was solved numerically. The eigenfunctions for the given azimuthal wave number $m$ were expanded in the series of Legendre polynomials,
\begin{equation}
    S =\ \mathrm{e}^{\mathrm{i}(m\phi - \omega t)}\sum\limits_{l=\max(\mid m\mid , 1)}^K S_l P_l^{\mid m\mid }(\mu ) ,
    \label{19}
\end{equation}
and similarly for $V$ and $W$. This gives the matrix equation for coefficients $S_l$, $V_l$ and $W_l$ of the expansion. The expansion (\ref{19}) with 100 harmonix ($K = \max(\mid m\mid , 1) +99$)  provides sufficient numerical resolution. The actual number of equations in the system is not about $3K$ but half as many because the complete system of equations splits into two independent subsystems governing modes of different types of equatorial symmetry (cf. Section \ref{symmetry} below). Computations were performed for the azimuthal wave numbers $\mid m \mid \leq 10$. Modes of higher $m$ are a matter of local theory.

Two types of numerical codes were used. One of them employs a standard routine of the {\sl EISPACK} library to compute the eigenvalue of the most rapidly growing mode, i.e. the largest $\Im ({\hat\omega})$. Another type of solvers employed the inverse iteration method (also known as the \lq\lq fixed-point method") to find the eigenvalue closest to a seed eigenvalue and the corresponding eigenmode. This method is convenient for following the dependence of an eigenvalue on the problem parameters as well as for finding the structure of the eigenmode of a given eigenvalue.

Computations were performed with the following values of the dissipation parameters of Eq.\,(\ref{15}),
\begin{equation}
    \epsilon_\chi = 10^{-4},\ \ \epsilon_\nu = 2\times 10^{-10} ,
    \label{20}
\end{equation}
characteristic of the upper radiation zone of the sun.
Dependence on thermal conductivity has been discussed in K13.
With the dissipation parameters fixed, only two controlling parameters remain variable: the normalized radial scale $\hat{\lambda}$ of Eq.\,(\ref{13}) and rotational shear $Q$ of Eq.\,(\ref{18}).
\subsection{Symmetry Properties}\label{symmetry}
As is usual for global stability problems in spherical geometry \citep{RKH13}, there are two types of modes of different equatorial symmetry. Symmetric modes have the entropy disturbances and also $u_r$ and $u_\phi$ components of velocity perturbations symmetric about the equator. $u_\theta$ is antisymmetric for these modes. In terms of the flow potentials, this reads:  $V(\mu ) = V(-\mu )$, $W(\mu ) = -W(-\mu )$ and $S(\mu ) = S(-\mu )$. The symmetry can be understood as being defined relative to the mirror-reflection about the equatorial plane. The notations Sm and Am will be used for symmetric and anti-symmetric modes, respectively, where $m$ is the azimuthal wave number. The antisymmetric modes with $V(\mu ) = -V(-\mu )$, $W(\mu ) = W(-\mu )$ and $S(\mu ) = -S(-\mu )$ have entropy disturbances as well as $u_r$ and $u_\phi$ antisymmetric about the equator but symmetric $u_\theta$.

A more significant property of the system of equations (\ref{11}), (\ref{12}) and (\ref{17}) is its symmetry relative to the transformation
\begin{equation}
    (q,m,\hat\omega ,W,V,S) \rightarrow (-q,-m,-\hat\omega^*,W^*,-V^*,S^*) ,
    \label{21}
\end{equation}
where the upper star means a complex conjugate. The symmetry rule in particular means that every mode with an azimuthal wave number $m$ growing for rotational shear $q$ has a counterpart mode for the shear of opposite sense, $-q$, growing at the same rate and having the azimuthal wave number $-m$. Therefore, if the results for a certain sign of $q$ are known, they also define the results for the rotational shear of opposite sense. Subrotation ($q > 0$) is expected for differential rotation caused by spindown of solar-type stars. The sign of $q$ is, however,  uncertain when differential rotation results from non-uniform compression/expansion in evolving stars. This paper reports the results for positive $q$, i.e., for angular velocity increasing inwards.

It may be noted that the symmetry rule of Eq.\,(\ref{21}) can be generalized to an arbitrary - not necessarily radial - differential rotation. In this case, the transformation $q\rightarrow -q$ in Eq.\,(\ref{21}) should be replaced by $\partial\Omega /\partial z \rightarrow -\partial\Omega /\partial z$ keeping the latitudinal differential rotation unchanged, i.e., the radial shear should be transformed as follows: $\mu r \partial \Omega /\partial r \rightarrow - \mu r \partial \Omega /\partial r - 2(1 -\mu^2)\partial\Omega/\partial\mu$.

The presence of the parameter $q$ of the background equilibrium in Eq.\,(\ref{21}) means that stability properties depend on the sign of the azimuthal wave number $m$: stability for given $q$ depends on the sign of $m$.
Such a dependence means that a certain handedness is inherent to the system and the unstable modes can possess helicity \citep{RKE12}. The absolute value of helicity is indefinite in the linear stability problem. The relative kinetic helicity,
\begin{equation}
    H_\mathrm{rel} = \langle {\vec u}\cdot ({\vec\nabla}\times{\vec u})\rangle /(k\overline{u^2}) ,
    \label{22}
\end{equation}
can, however, be defined. The angular brackets in this equation mean the azimuthal averaging:
\begin{equation}
    \langle {\vec u}\cdot ({\vec\nabla}\times{\vec u})\rangle =
    \frac{1}{2\pi}\int\limits_0^{2\pi} {\vec u}\cdot ({\vec\nabla}\times{\vec u}) \mathrm{d}\phi .
    \label{23}
\end{equation}
For axisymmetric modes, it can be understood as phase averaging (linear solutions are defined up to the phase multiplier $\mathrm{e}^{\mathrm{i}\phi}$). The overline in Eq.\,(\ref{23}) means the full-sphere averaging:
\begin{equation}
    \overline{u^2} = \frac{1}{2}\int\limits_{-1}^1\langle u^2\rangle \mathrm{d}\mu .
    \label{24}
\end{equation}
The relative helicity (\ref{22}) written in terms of the flow potentials reads
\begin{eqnarray}
    H_\mathrm{rel} &=& \frac{1}{\sin^2\theta\ \overline{v^2}}
    \Bigg[\Re\left( mW +\sin\theta\frac{\partial V}{\partial\theta}\right)
    \Re\left( mV +\sin\theta\frac{\partial W}{\partial\theta}\right)
    \nonumber \\
     &+&\Im\left( mW +\sin\theta\frac{\partial V}{\partial\theta}\right)
    \Im\left( mV +\sin\theta\frac{\partial W}{\partial\theta}\right)\Bigg] ,
    \label{25}
\end{eqnarray}
where $\overline{v^2} = \overline{u^2}/(\Omega_0^2r^2)$ is the normalized kinetic energy. Eq.\,(\ref{25}) shows that only axisymmetric modes have to combine both toroidal and poloidal flows to possess helicity. Non-axisymmetric toroidal or poloidal flows alone can be helical.
\subsection{Two Modes of Stable Oscillations}
Solutions for particular limiting cases are helpful in interpreting the results of subsequent computations. Two families of stable oscillations can be found in the case of zero dissipation ($\nu = \chi = 0$).

1. In non-rotating fluid ($\hat\Omega = 0$), the eigenvalue equations simplify to read
\begin{equation}
    \hat{\omega}\hat{L}V = -{\hat\lambda}^2\hat{L}S ,\ \ \
    \hat{\omega} S = \hat{L}V ,\ \ \ \hat{\omega}\hat{L} W = 0.
    \label{26}
\end{equation}
Apart from the trivial solution of steady toroidal vortices ($\hat\omega = S = V = 0$ and $W$ is an arbitrary function), there is a family of poloidal ($W = 0$) oscillations:
\begin{equation}
    \hat\omega = \pm \hat\lambda \sqrt{l(l+1)},\ \ \ l =1,2,...
    \label{27}
\end{equation}
The spectrum (\ref{27}) depends on $\Omega_0$ due to normalization only. The expression for dimensional frequencies,
\begin{equation}
    \omega = \pm\frac{N}{kr}\sqrt{l(l+1)} ,\ \ \ l =1,2,...
    \label{28}
\end{equation}
shows that these are the internal gravity waves, or $g$-modes.

2. Another family of low-frequency ($\omega \sim \Omega$) oscillations can be found for uniform rotation ($q = 0$) and strongly subadiabatic stratification ($\hat\lambda \gg 1$ or $N \gg \Omega kr$). Equation (\ref{12}) in the leading order in $\hat\lambda$ gives $S = 0$. Then, Eq.\,(\ref{17}) yields $V = 0$, and Eq.\,(\ref{11}) provides the spectrum of purely toroidal oscillations
\begin{eqnarray}
    &&\hat{\omega} = m{\hat\Omega}\left( 1 - \frac{2}{l(l+1)}\right) ,
    \\
    \label{29}
    &&m=\pm 1, \pm 2, ...,\
    l = \mid m\mid , \mid m\mid + 1, \mid m\mid + 2, ...
    \nonumber
\end{eqnarray}
These are $r$-modes or Rossby waves - global vortices drifting in the co-rotating frame in counter-rotation direction.
\section{RESULTS AND DISCUSSION}\label{RD}
\subsection{Baroclinic Instability as Stability Loss to $g$- and $r$-Modes Excitation}
All the unstable modes are oscillatory. The overstability was interpreted as stability loss to excitation of $g$- and $r$-modes in K13.
Now we confirm this guess by computations. The computations show that the $g$- and $r$-modes transform continuously into modes of baroclinic instability as the differential rotation is varied smoothly from zero to a sufficiently large value.

\begin{figure*}
    \begin{center}
    \includegraphics[width=7.0 cm, height = 7.5 cm]{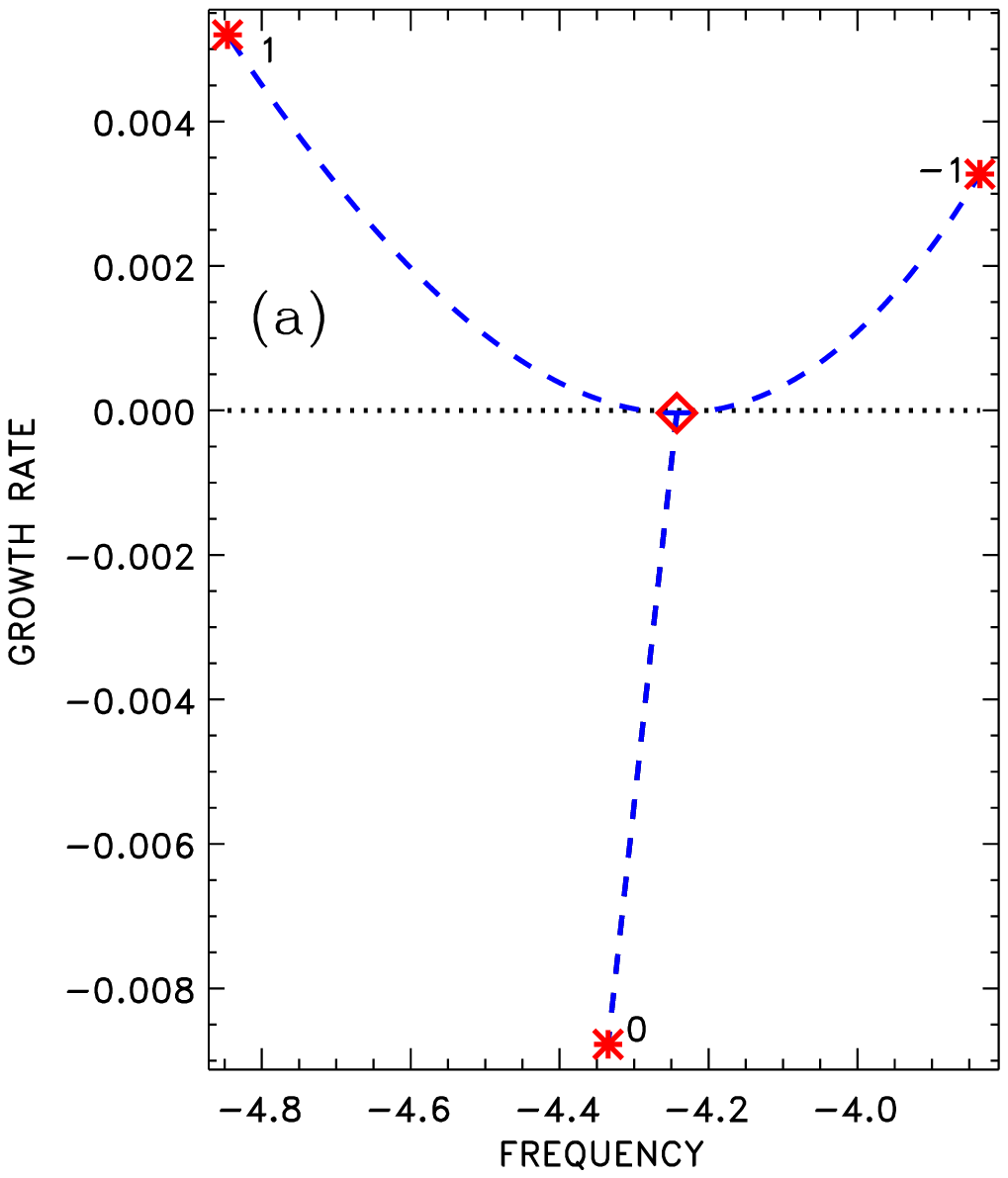}
    \hspace{0.4 cm}
    \includegraphics[width=7.0 cm, height = 7.5 cm]{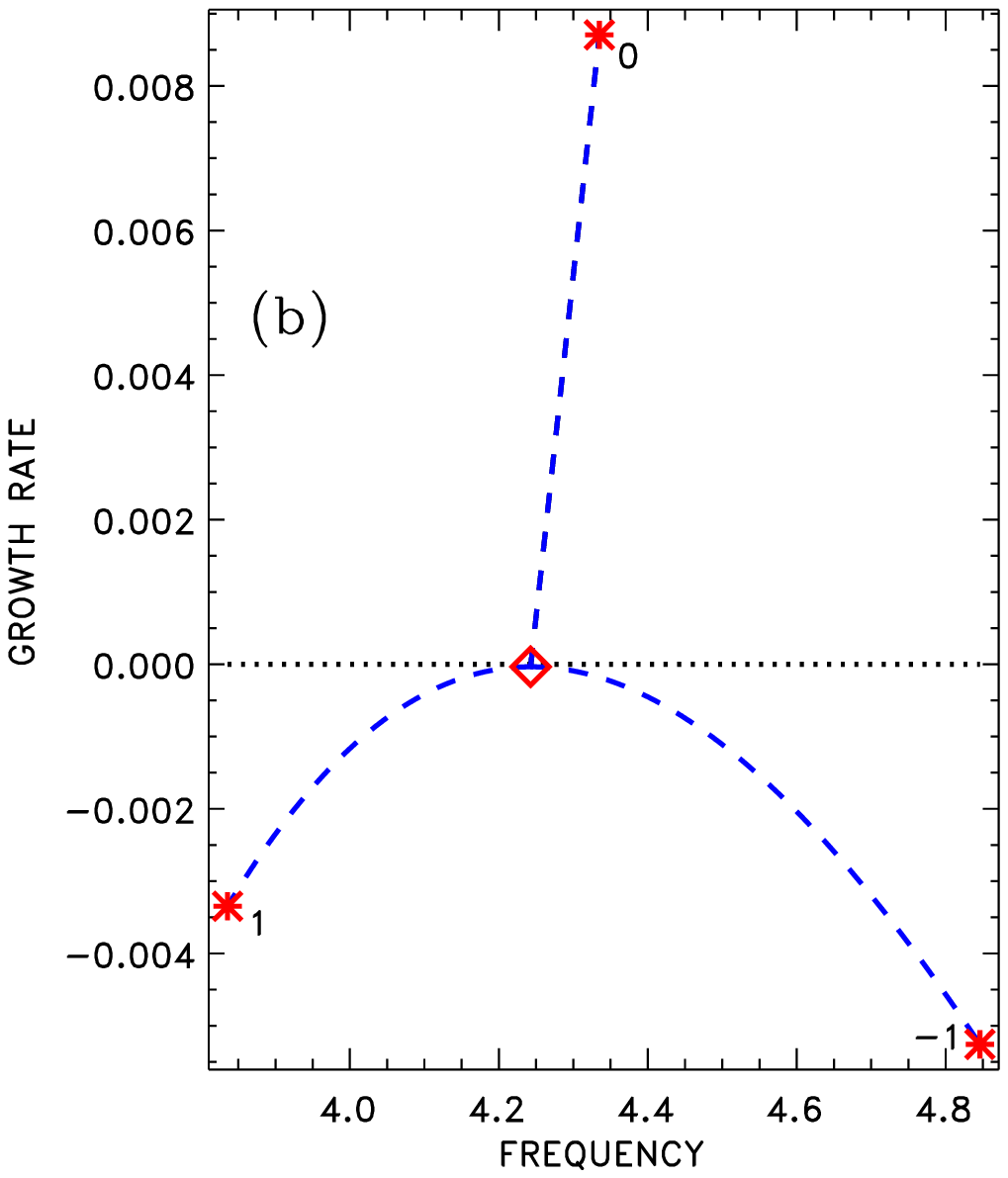}
    \\[0.3 cm]
    \includegraphics[width=7.0 cm, height = 7.5 cm]{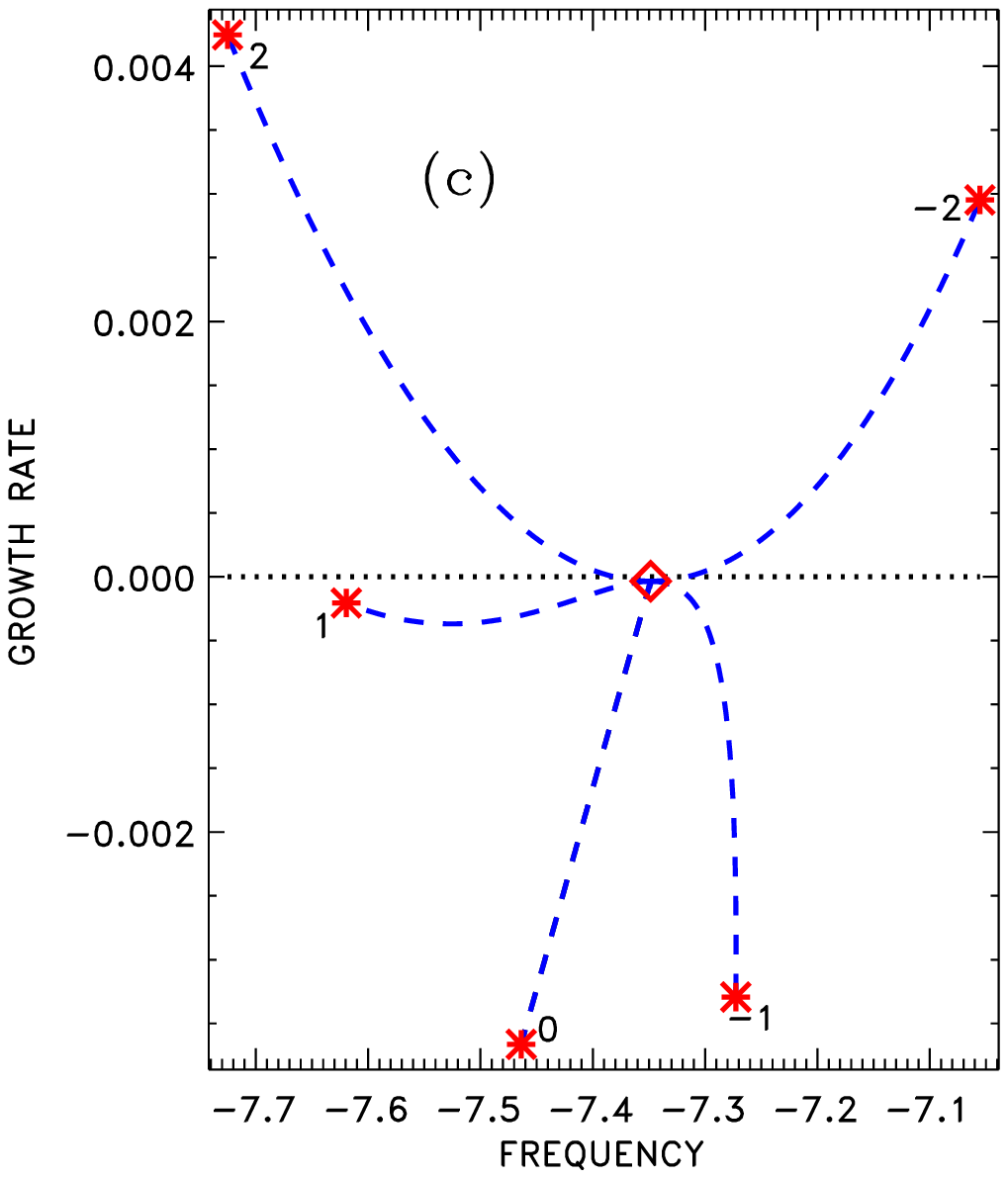}
    \hspace{0.4 cm}
    \includegraphics[width=7.0 cm, height = 7.5 cm]{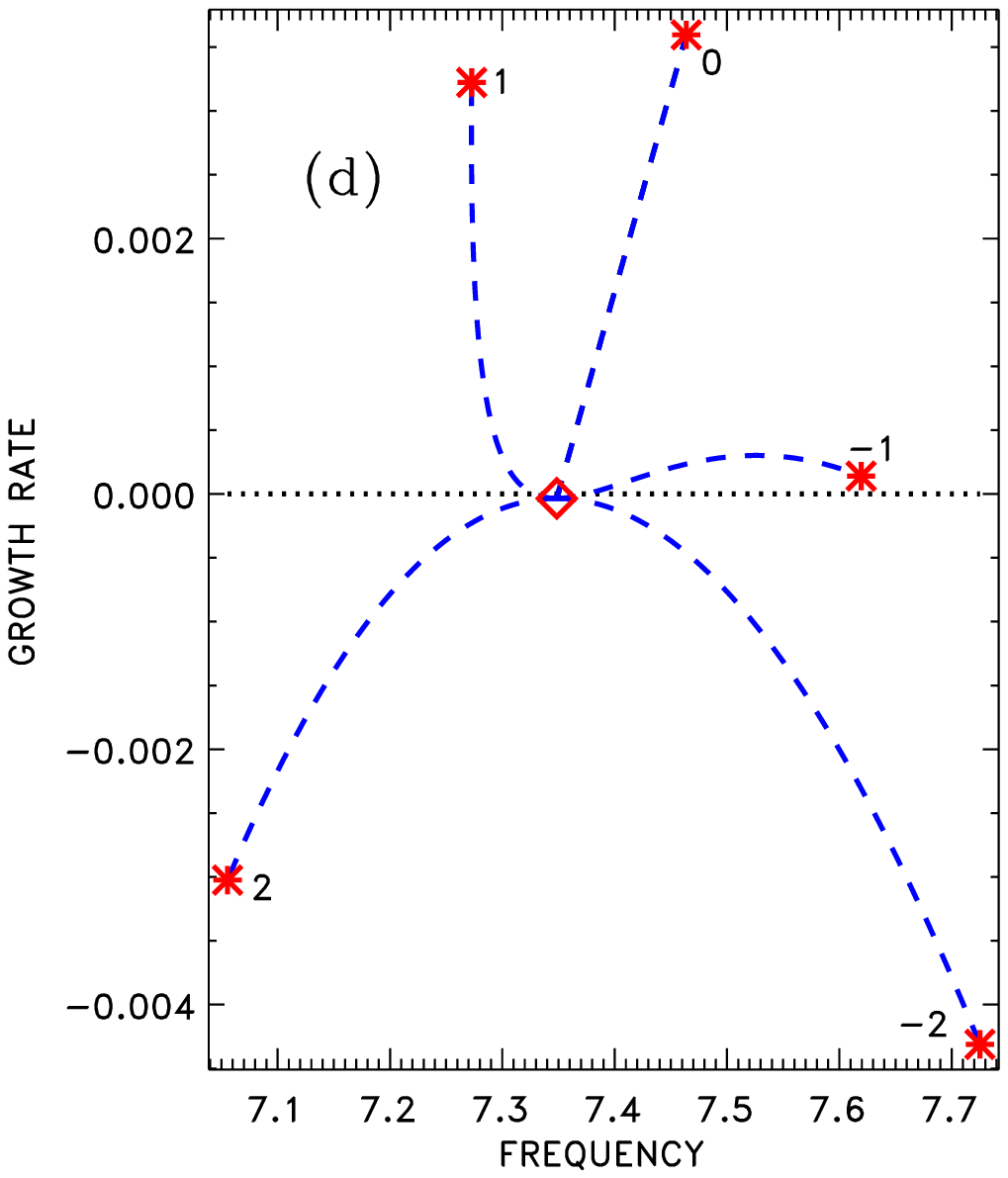}
    \end{center}
    \caption{Diamonds show the eigenvalues of $g$-modes of Eq.\,(\ref{27}) for $l=1$ (a,b) and $l=2$
        (c,d). These eigenvalues for non-rotating fluid ($\hat\Omega = 0$) do not depend on $m$. As the rotation rate increases, the eigenvalues follow the trajectories shown by dashed lines. The final positions for $\hat\Omega = 1$ are shown by asterisks. The positions are marked by corresponding $m$. The normalized shear (\ref{18}) and radial wave length (\ref{13}) were fixed to $Q = 0.01$ and $\hat\lambda = 3$.
    }
    \label{f2}
\end{figure*}

In case of $g$-modes, the shear parameter $Q$ of Eq.\,(\ref{18}) and $\hat\lambda$ of Eq.\,(\ref{13}) were fixed and the normalized rotation rate $\hat\Omega$ was varied from zero to one. Figure\,\ref{f2} shows trajectories of several eigenvalues, which initially (for $\hat\Omega = 0$) belong to $g$-modes of Eq.\,(\ref{27}), on the plane of growth rate $\gamma = 2\pi\Im (\hat\omega )$ and frequency $\Re (\hat\omega ) - m{\hat\Omega}$ in co-rotating frame. In non-rotating fluid ($\hat\Omega =0$), the frequencies do not depend on the azimuthal wave number $m$. The \lq\lq initial" eigenvalues have minor decay rates $\gamma < 0$ due to the finite diffusion of Eq.\,(\ref{20}). Rotation brakes the degeneracy in the azimuthal wave number because of the Coriolis acceleration.  As rotation rate grows, the modes attain considerable growth or decay rates dependent on $m$. Their frequencies are changed mildly by rotation. The velocity field of these modes is dominated by the poloidal flow. The kinetic energy of the disturbances is the sum of kinetic energies of their poloidal and toroidal parts,
\begin{equation}
    \overline{v^2} = \overline{v^2_\mathrm{p}} + \overline{v^2_\mathrm{t}} = \frac{1}{4}\sum\limits_l l(l+1) \left( \mid V_l\mid^2 + \mid W_l\mid^2\right) .
    \label{30}
\end{equation}
It is $\overline{v^2_\mathrm{p}} \gg \overline{v^2_\mathrm{t}}$ for all modes of Fig.\,\ref{f2}. The unstable modes in rotating fluid remain close though not identical to the original $g$-modes. The name \lq\lq $g$-modes" is kept for these rotationally modified almost poloidal oscillations.

All but one mode of Fig.\,\ref{f2} growing at $\hat\Omega = 1$ are the most rapidly growing modes of their symmetry type. E.g., the axisymmetric mode of part (b) of this Figure is the most rapidly growing A0 mode and the mode with $m=2$ of part (c) is the most rapidly growing S2 mode of baroclinic instability. There only exception is the $m=-1$ mode (A-1) of part (d). The most rapidly growing A-1 is an $r$-mode.

\begin{figure}[htb]
    \begin{center}
    \resizebox{\hsize}{!}{\includegraphics{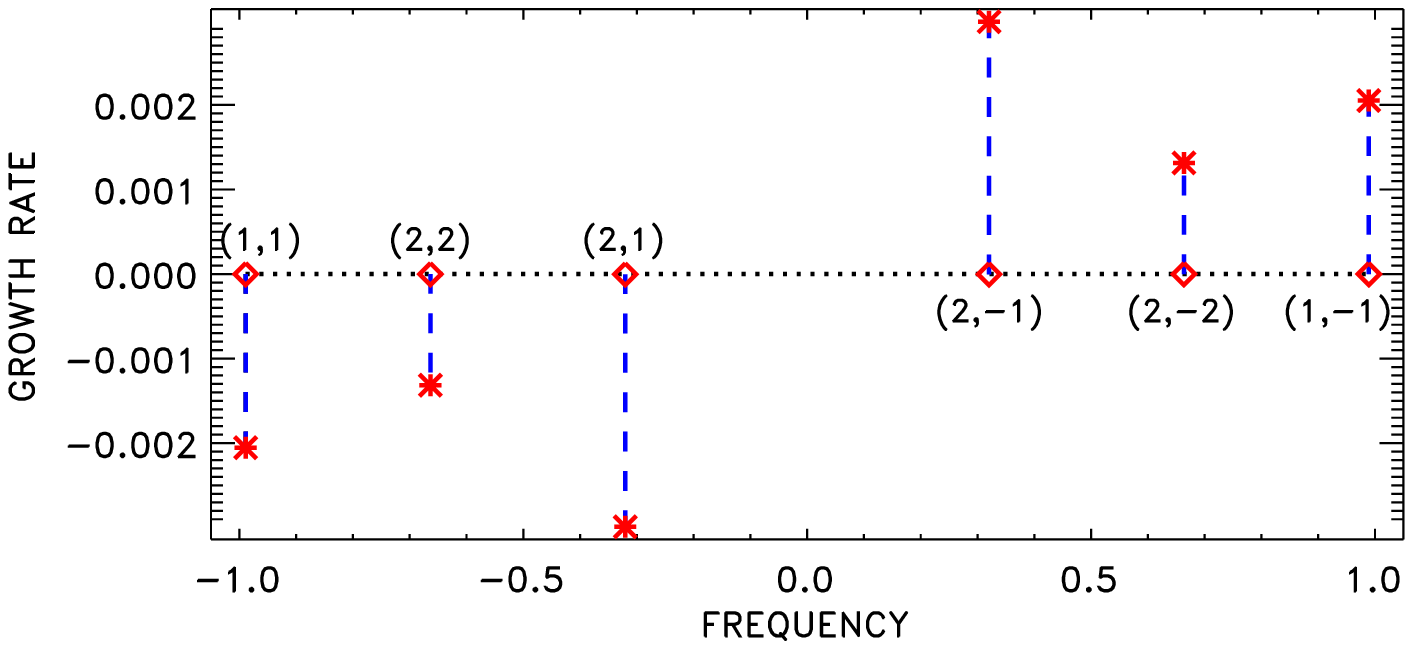}}
    \end{center}
    \caption{Diamonds show eigenfrequencies of several $r$-modes of Eq.\,(\ref{29})
        in co-rotating frame, $\hat{\omega} - m$. The frequencies are marked by the corresponding meridional and azimuthal wave numbers $(l,m)$. As differential rotation grows from zero to $Q=0.01$, the eigenvalues follow trajectories shown by the dashed lines. The modes with negative $m$ become unstable. Asterisks show the eigenvalues for $Q=0.01$. $\hat\lambda =3$.
    }
    \label{f3}
\end{figure}

A similar plot for $r$-modes is shown in Fig.\,\ref{f3}. These modes do not exist without rotation. In this case, therefore, rotation rate is fixed, ${\hat\Omega} = 1$, and differential rotation increased gradually from zero to a finite value. In the case of uniform rotation, the $r$-modes with positive and negative $m$ are physically identical. Differential rotation brakes the identity so that the modes with positive $m$ decay and those with negative $m$ grow exponentially. Drift rates (frequencies) are not changed by shellular differential rotation. The flow in unstable disturbances of Fig.\,\ref{f3} is dominated by its toroidal part, $\overline{v^2_\mathrm{t}} \gg \overline{v^2_\mathrm{p}}$. The disturbances remain close though not identical to the Rossby vortices. The name \lq\lq $r$-modes" is, therefore, kept for these low-frequency almost toroidal eigenmodes.

It may be noted that the correlation of entropy and radial velocity,
\begin{equation}
    G = \overline{S u_r} / \sqrt{\overline{S^2}\,\overline{u_r^2}}\ ,
    \label{31}
\end{equation}
is positive for all growing modes. For decaying disturbances, the correlation is predominantly negative (it can be $G > 0$ for the slowly  decaying modes very close to the instability border). The instability is, therefore, fed by release of gravitational energy, as it should be the case for the baroclinic instability (Fig.\,\ref{f1}).
\subsection{Stability Maps}
\begin{figure}
    \begin{center}
    \resizebox{\hsize}{!}{\includegraphics{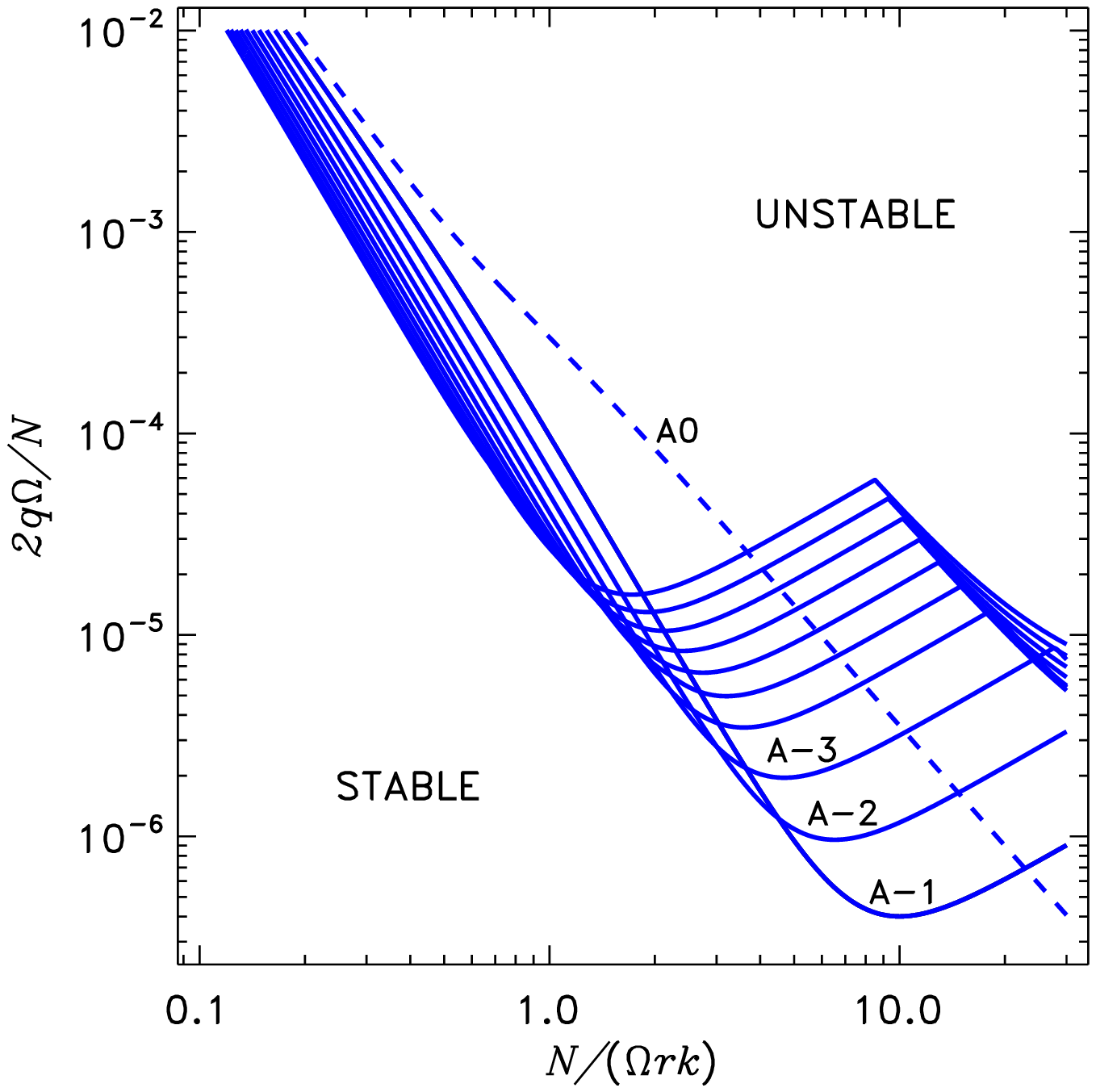}}
    \end{center}
    \caption{Marginal stability lines on the plane of normalized differential rotation
        (\ref{18}) and radial wave length (\ref{13}). The lines are marked by the corresponding symmetry notations. Only those modes which require the smallest differential rotation for their excitation in some range of radial scales are shown.
    }
    \label{f4}
\end{figure}

Figure~\ref{f4} shows the lines separating regions of stability and instability on the plane of two basic parameters measuring differential rotation and radial scale of disturbances. Different lines correspond to the modes of different equatorial and axial symmetries. The marginal stability lines are shown only for those modes, which require the smallest differential rotation for excitation in some range of radial wave lengths. As the wave length decreases, modes with increasingly larger $\mid m\mid$ have the lowest threshold value of differential rotation for the onset of the instability.

It is $N/\Omega > 300$ in the upper radiation zone of the sun. Accordingly, the computations were performed for ${\hat\lambda} \leq 30$ so that $kr \geq 10$ remains at least moderately large. The dashed line in Fig.\,\ref{f4} represents the $g$-mode A0. The full lines in the regions of their minima belong to $r$-modes. Kinks in the full lines signify transitions from $r$- to $g$-modes with increasing radial wave length. The $g$-modes dominate on the relatively long radial wave-lengths side of the Figure, and $r$-modes - at relatively short wave lengths.

Even a very small differential rotation with $Q \sim 10^{-6}$ can provoke instability. With $N/\Omega \sim 10^3$ this corresponds to very small rotational shear of $q \sim 10^{-3}$. The critical shear decreases with increasing rotation rate.

\begin{figure}
    \begin{center}
    \resizebox{\hsize}{!}{\includegraphics{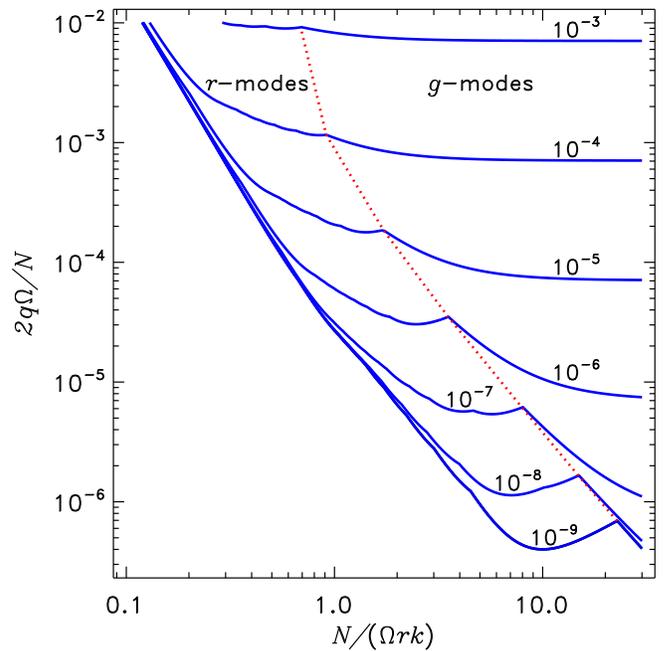}}
    \end{center}
    \caption{Lines of constant growth rates on the plane of normalized differential rotation
            $Q$ (\ref{18}) and radial wave length $\hat\lambda$ (\ref{13}). The isolines show the maximum growth rates among unstable modes. The dotted line is the watershed between the regions where $r$- or $g$-modes are growing most rapidly.
    }
    \label{f5}
\end{figure}

Figure\,\ref{f5} shows the lines of constant growth rate $\gamma = 2\pi\Im ({\hat\omega})$. The growth rates are normalized so that the amplitude of excitation increases by a factor of $\mathrm{e}^\gamma$ in one revolution of a star. The Figure shows maximum growth rates among unstable modes. This maximum belongs to $r$- or $g$-modes depending on the position on the plot.
The dotted line on the stability map shows the border between the regions where $r$- or $g$-modes grow most rapidly.
The wavy shape of the lines at relatively small $\hat\lambda$ is caused by the switching of the maximum growth rates to increasingly high azimuthal wave numbers as $\hat\lambda$ decreases (Fig.\,\ref{f4}).

The growth rates are small. The e-folding times are, however, short compared to time scales of stellar evolution. Even in a slowly rotating star like the sun, the e-folding time for a radial differential rotation of 1\% is about 10\,000 years. This time decreases in proportion to $\Omega^{-2}$ in faster rotators.
The e-folding time may be as short as 1 year in the infant sun arriving on the main sequence with a rotation period of about 1 day and having appreciable differential rotation in its radiation core yet decoupled from convective envelope \citep{HN87}.

Our computations do not help to decide which modes are preferred by baroclinic instability. The $r$- and $g$-modes have comparable growth rates. The threshold differential rotation for onset of $g$-modes instability decreases with increasing $\hat\lambda$. The assumption of short radial scales does not allow us to see whether the decrease saturates for sufficiently large radial scales. The analysis should be global not only in horizontal dimensions but in radius also to decide about this. Here we  meet a rare case where instability in the subadiabatically stratified radiation zone can be global in radius (cf., however, \citet{BU13a,BU13b}).
\subsection{Instability Patterns and Kinetic Helicity}
\begin{figure*}
    \begin{center}
    \includegraphics[width=7.9 cm]{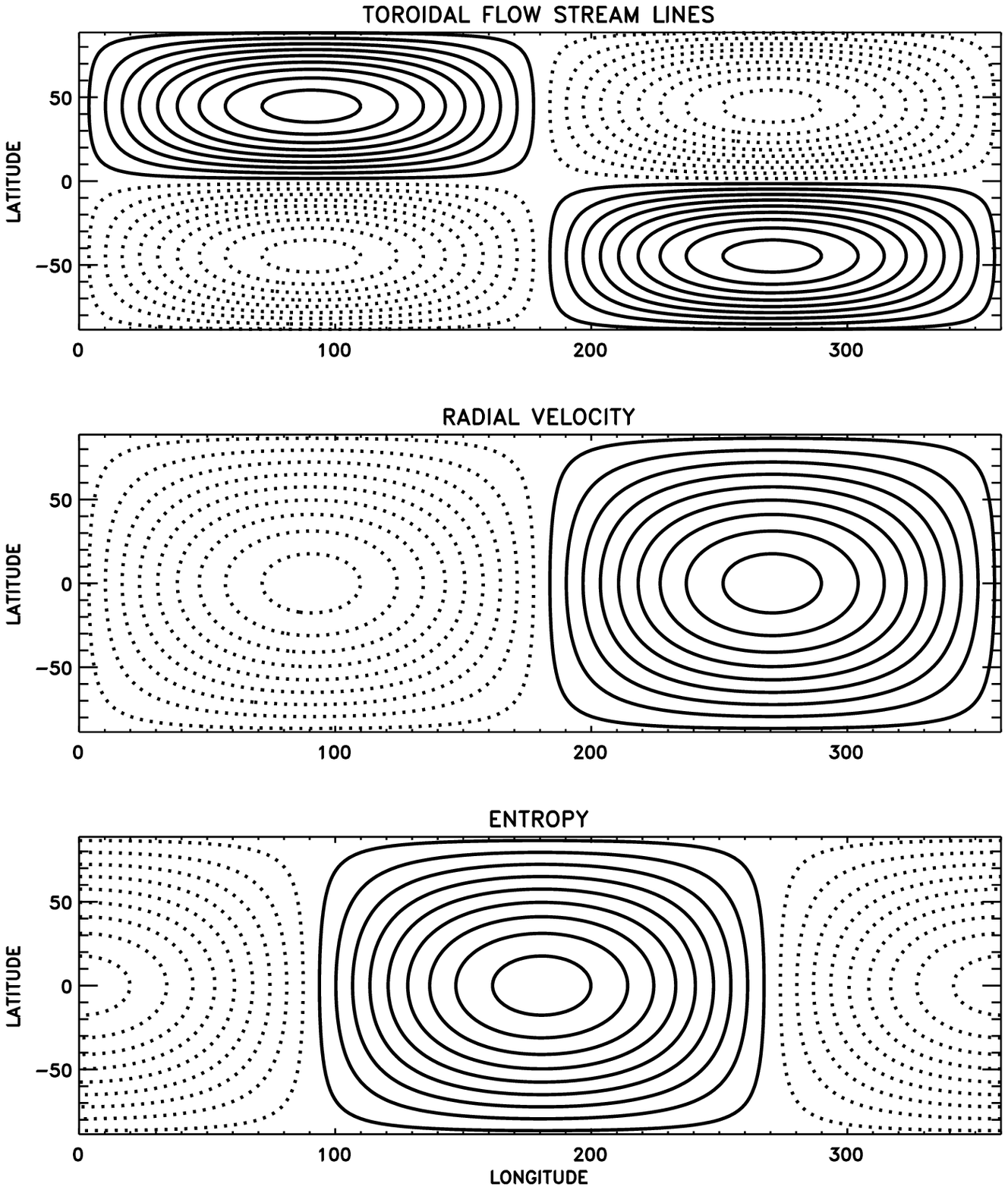}
    \hspace{0.7 cm}
    \includegraphics[width=7.9 cm]{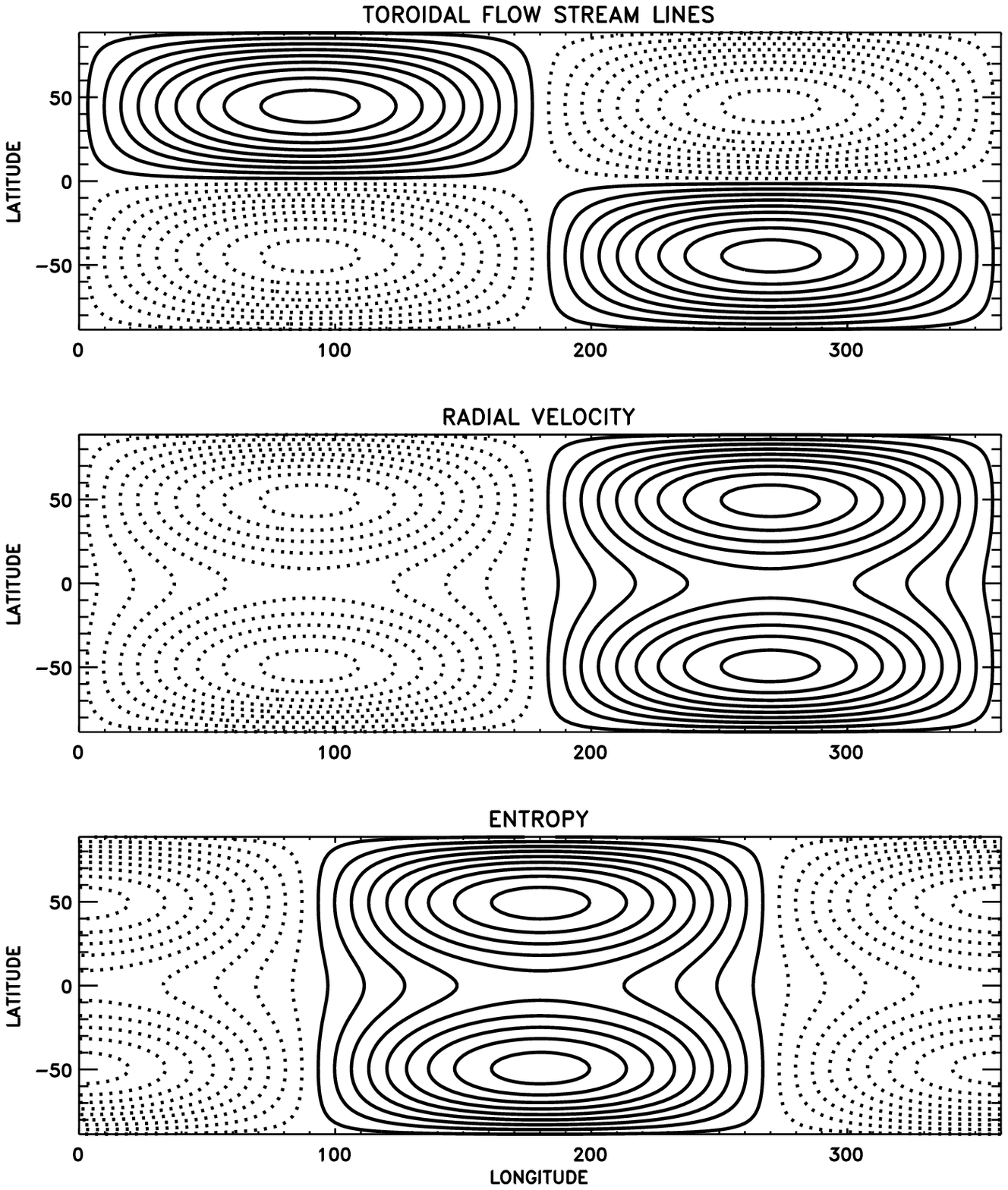}
    \end{center}
    \caption{{\sl Left panel:} Stream-lines of toroidal flow, iso-contours of radial velocity and entropy disturbances for unstable S1 mode. Full (dotted) lines show clockwise (anti-clockwise) circulation for toroidal flow and positive (negative) levels for radial velocity and entropy.
    This is  $g$-mode with $\overline{v_\mathrm{p}^2}/\overline{v_\mathrm{t}^2} = 34.6$ and growth rate $\hat{\gamma} = 4.93\times 10^{-4}$. {\sl Right panel:} The same for S-1 mode, which is $r$-mode with $\overline{v_\mathrm{p}^2}/\overline{v_\mathrm{t}^2} = 1.93\times 10^{-4}$ and growth rate $\hat{\gamma} = 2.94\times 10^{-4}$. The modes were computed for $Q=10^{-3}$ and $\hat{\lambda} = 3$.
    }
    \label{f7}
\end{figure*}

Figure\,\ref{f7} shows patterns of flow and entropy disturbances for unstable S1 and S-1 modes, which are $g$- and $r$-mode, respectively. The patterns look similar in spite of the difference in the modes nature. Radial velocity and entropy patterns in both modes are phase-shifted so that the sign of their correlation $G$ of Eq.\,(\ref{31}) cannot be estimated by eye. The correlation is nevertheless positive, as it should be for baroclinic instability:  $G = 2.17\times 10^{-5}$ and $G = 3.67\times 10^{-3}$ for $g$- and $r$-mode, respectively. The smallness of the correlation explains the small growth rates. The larger correlation for $r$-mode does not lead to larger growth rate because the poloidal part of the flow in this mode is small.

\begin{figure}[thb]
    \begin{center}
    \resizebox{\hsize}{!}{\includegraphics{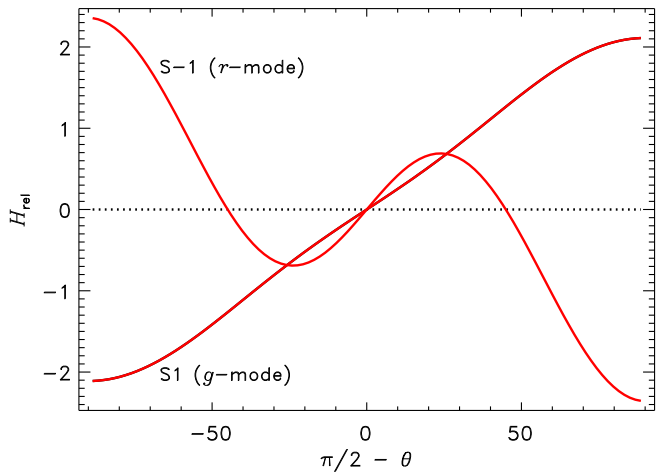}}
    \end{center}
    \caption{Latitudinal profiles of relative helicity (\ref{25}) for two unstable modes of Fig.\,\ref{f7}.
    }
    \label{f8}
\end{figure}

This Figure as well as Figs.~\ref{f2} and \ref{f3} shows that the stability properties depend on the sign of the azimuthal wave number $m$. This means that baroclinic rotating fluids possess certain handedness. The unstable modes can, therefore, be expected to possess helicity.
Figure\,\ref{f8} shows that both modes of Fig.\,\ref{f7} are indeed helical. The relative helicity is not small. It is mainly contributed to by the scalar product of horizontal velocity and vorticity. Radial velocity and vorticity are both small for these modes, which are horizontally global but short-scaled in radius. The helicity profiles are anti-symmetric about the equator as it should be the case with an effect of Coriolis force. Helicity of unstable $g$-modes is predominantly positive in the northern hemisphere (K13). Latitudinal profiles of $r$-modes helicity are less regular with the tendency to be negative in the northern hemisphere. Both types of modes, however, show global helicity patterns.

Kinetic helicity (of either sign) is known to be important for dynamos \citep{BS05}. Baroclinic instability may, therefore, have some bearing on the stellar radiation zone's magnetism. Solar-type stars have appreciable differential rotation in their radiation cores in about initial 100 Myr of their main-sequence life \citep{Pea89}. This time is long compared to characteristic growth time of the instability. It suffices for the helical baroclinic modes to generate internal magnetic fields in the young stars, similar to helical convection of their envelopes.

Radiation zones of solar-type stars are deep beneath the outer convective envelopes and not accessible to observation. More massive stars have external radiation zones. Differential rotation can be present in young stars of this type due to non-uniform contraction to the main sequence.
\citet{Aea13} observed a rapid (several years) change in a magnetic topology of the almost entirely radiative pre-main sequence star HD~190073. They interpreted this change as a manifestation of the start of a dynamo operation in a newly born convective core. An alternative interpretation could be a \lq\lq baroclinic dynamo" in the radiative envelope.
Observations of \citet{Hea09} hint at a decline in magnetic fields in the Herbig Ae/Be stars with age. They interpreted the decline as indication of a dynamo mechanism that decays with age. In relation to the possibility of dynamo by baroclinic instability, it is tempting to know whether $g$- and $r$-modes of global oscillations presumably excited by the instability can be observed on Herbig stars.

Another widely discussed possibility for magnetic field generation in radiation zones is the Tayler-Spruit dynamo driven jointly by differential rotation and instability of the toroidal magnetic field \citep{S02}. The difference with baroclinic instability, however, is that \citet{T73} instability of the toroidal field does not produce net helicity. More specifically, the instability properties are symmetric relative to the change of sign of the azimuthal wave number: the unstable modes with $m$ of opposite sign have equal growth rates and helicities of opposite sense \citep{RKE12} so that a fluctuating helicity with zero mean can be expected. Direct numerical simulations of Tayler instability by \citet{ZBM07} did not show a dynamo.
\section{CONCLUSIONS}\label{Con}
Linear analysis does not permit determination of the eventual state to which instability leads. It may be noted, however, that the correlation of radial velocity and entropy was positive for all unstable modes in our computations. The instability, therefore, reduces baroclinicity to produce slight deviation from the thermal wind balance of Eq.\,(\ref{5}). Any such deviation results in a (weak) meridional flow that reacts back on stratification and differential rotation to re-establish the balance. In this way, the instability can indirectly affect the differential rotation. Another possibility is the angular momentum transport by turbulence resulting from the instability. Turbulence can be expected in view of the vast variety of baroclinic instability modes. Turbulence in radiation zones, irrespective of its origin, is highly anisotropic with predominance of horizontal motions, $\overline{u^2_r}/\overline{u^2} \sim \Omega^2/(\tau^2N^4) \ll 1$, where $\tau$ is the turbulence correlation time \citep{KB12}. Turbulence can transport angular momentum to reduce the differential rotation. This transport is not by the eddy viscosity only. Anisotropic turbulence produces non-diffusive fluxes of angular momentum \citep{R89} so that redistribution of angular momentum proceeds much faster than eddy diffusion of chemical species. Since the threshold value of differential rotation for the onset of baroclinic instability is very low (Fig.\,\ref{f4}), the instability may lead to an almost uniform rotation.

Baroclinic instability may be relevant to the origin of magnetic fields in radiation zones of stars. Convective instability in rotating stars is known to be capable of generating magnetic fields. The kinetic helicity of rotating convection plays a key role in the dynamo process. Baroclinic instability is helical as well (Fig.\,\ref{f8}). It can drive a (transient) dynamo in the radiation zone until the differential rotation declines.

The instability excites two families of physically different modes - the Rossby waves and internal gravity waves. Judging from Fig.\,\ref{f5},  $g$-modes have slightly larger growth rates for given rotational shear compared to $r$-modes. If $g$-modes are preferentially excited, helicity pattern positive in the northern hemisphere and anti-symmetric about the equator can be expected from the instability (Fig.\,\ref{f8}).

Differential rotation is not the only possible cause of baroclinicity. The influence of a magnetic field is another possibility \citep{MZ05,SBZ11}. Equation (\ref{3}) for the background equilibrium modifies taking into account the axisymmetric toroidal field, $B$, to read
\begin{equation}
    r\sin\theta\frac{\partial (\Omega_\mathrm{A}^2 - \Omega^2)}{\partial z} =\
    \frac{1}{\rho^2}\left( {\vec\nabla}\rho\times{\vec\nabla}P^*\right)_\phi ,
    \label{33}
\end{equation}
where $\Omega_\mathrm{A} = B/(\sqrt{4\pi\rho}\ r\sin\theta)$ is the angular Alfv\'en frequency, and $P^* = P + B^2/(8\pi)$ is the total pressure. Not only non-conservative centrifugal force but also magnetic tension can produce baroclinicity. This raises the question of whether magnetically induced baroclinic instability is possible. The question is complicated by the presence of another magnetic instability \citep{T73}. However, \citet{BU12} took the influence of the toroidal magnetic field on stratification into account to find that the field is always unstable irrespective of whether criteria for \citet{T73} instability are satisfied.

Baroclinic instability has a bearing not only on stars. Already \citet{TT83} pointed to this instability as a possible source of turbulence in accretion disks. More recently, this possibility was studied by \citet{KB03} and \citet{NGU13}. The stratorotational instability of a Couette flow \citep{SR05} is most likely of baroclinic type as well.

Computations in this paper were performed for radial differential rotation. Weak meridional flow in early-type stars leads to rotation laws dependent on latitude \citep{ER13}. The study of baroclinic instability for differential rotation including dependence on latitude can, therefore, be noted as a perspective for future work.
\acknowledgments
The author is thankful to the Russian Foundation for Basic Research
(projects 12-02-92691\_Ind, 13-02-00277) and to the Ministry of Education
and Science of the Russian Federation (contract 8407) for their support.
\bibliographystyle{apj}
\bibliography{kitbib}
\end{document}